\documentclass[%
 reprint,
showpacs,
 amsmath,amssymb,
 aps,
 pra,
longbibliography,
floatfix
]{revtex4-1}

\usepackage[utf8]{inputenc}	
\usepackage[T1]{fontenc}	
\usepackage[]{graphicx}		
\usepackage[export]{adjustbox}
\usepackage{color}
\usepackage[english]{babel}
\usepackage{hyperref}
\usepackage{url}
\usepackage{todonotes}
\usepackage{lipsum}
\usepackage{bm}

\usepackage{microtype}
\usepackage{braket}

\usepackage[position=top,caption=false]{subfig}

\let\originaleqref\eqref
\renewcommand{\eqref}{Eq.~\originaleqref}

\newcommand{\fref}[1]{\figurename~\ref{#1}}

\newcommand{\Tr}[1]{\mathrm{Tr}\left(#1\right)}
\newcommand{\tr}[1]{\mathrm{Tr}(#1)}

\newcommand{\C}[0]{\hat{c}}
\newcommand{\Cd}[0]{\hat{c}^\dagger}

\newcommand{\sx}[0]{\hat{\sigma}_\mathrm{x}}

\newcommand{\sz}[0]{\hat{\sigma}_\mathrm{z}}

\newcommand{\Qe}{Q_{\mathrm{e}}}
\newcommand{\rzo}{\rho_{01}}

\newcommand{\cjz}{\hat{c}_{0j}}
\newcommand{\cjo}{\hat{c}_{1j}}
\renewcommand{\H}{\hat{H}}

\newcommand{\D}{D_t}
\newcommand{\DT}{D_T}
\newcommand{\dD}{dD_t}
\newcommand{\rhoc}{\rho^{(\D)}}
\newcommand{\rhotc}{\tilde{\rho}^{(\D)}}
\newcommand{\rhon}{\rho^{(N_t)}}
\newcommand{\rhotn}{\tilde{\rho}^{(N_t)}}
\newcommand{\rhoy}{\rho^{(Y_t)}}
\newcommand{\rhoty}{\tilde{\rho}^{(Y_t)}}

\graphicspath{
{./figures/}
{/home/alexander/Dropbox/PhD/HypothesisTesting/Matlab/HypothesisTestingConditional/plots/paper/}
{/home/alexander/Dropbox/PhD/HypothesisTesting/Matlab/HypothesisTestingConditional/plots/homodyneCountingComparison/}
{/home/alexander/Dropbox/PhD/HypothesisTesting/figures/}
}

\begin{document}
\title{Hypothesis testing with a continuously monitored quantum system}
\author{Alexander Holm Kiilerich}
\email{kiilerich@phys.au.dk}
\author{Klaus Mølmer}
\email{moelmer@phys.au.dk}
\affiliation{Department of Physics and Astronomy, Aarhus University, Ny Munkegade 120, 8000 Aarhus C, Denmark}
\date{\today}

\bigskip

\begin{abstract}
In a Bayesian analysis, the likelihood that specific candidate parameters govern the evolution of a quantum system are conditioned on the outcome of measurements which, in turn, cause measurement backaction on the state of the system [M. Tsang, Phys. Rev. Lett. {\bf 108}, 170502 (2012)]. Specializing to the distinction of two candidate hypotheses, we study the achievements of continuous monitoring of the radiation emitted by a quantum system followed by an optimal projective measurement on its conditioned final state. Our study of the radiative decay of a driven two-level system shows an intricate interplay between the maximum information available from photon counting and homodyne detection and the final projective measurement on the emitter. We compare the results with theory predicting a lower bound for the probability to assign a wrong hypothesis by any combined measurement on the system and its radiative environment.
\end{abstract}

\maketitle
\noindent

\section{Introduction}
\label{sec:1}
Hypothesis testing is the task of assigning one of a discrete set of models to describe an observed system.
Measurements on quantum systems have random outcomes and the discrimination of two hypotheses $h_0$ and $h_1$ is a statistical inference problem. Viz. there is a probability $P(m=i|h_j)$ that measurement data processed to yield a binary outcome $m=0,1$ is (in)consistent with the true hypothesis $(i\neq j)i=j$
and a corresponding average probability that an erroneous hypothesis will be assigned,
\begin{align}\label{eq:Qe}
\Qe = P(m=1|h_0)P(h_0)+ P(m=0|h_1)P(h_1),
\end{align}
where $P(h_0)$ and $P(h_1)$ are the prior probabilities of each hypothesis.

Distinguishing two different Hamiltonians $\hat{H}_0$ and $\hat{H}_1$, governing the evolution of a closed quantum system, is achieved by discriminating the two quantum states $\rho_0(t) = \ket{\psi_0(t)}\bra{\psi_0(t)}$ (hypothesis $h_0$) and $\rho_1(t) = \ket{\psi_1(t)}\bra{\psi_1(t)}$  (hypothesis $h_1$), resulting from time evolution under each candidate Hamiltonian from a common initial state of the system.
Only orthogonal states can be discriminated unambiguously while, in general, the overlap between the candidate states defines a minimum error probability for any measurement protocol,
$\Qe\geq \Qe^{(\text{min})}$ where \cite{Helstrom1969}
\begin{align}\label{eq:QeMinPure}
\Qe^{(\text{min})} = \frac{1}{2}\left(1-\sqrt{1-4P(h_0)P(h_1)|\bra{\psi_0}\psi_1\rangle|^2}\right).
\end{align}
As derived by Helstrom \cite{Helstrom1969}, this bound can be saturated by performing a projective measurement of the operator
\begin{align}\label{eq:PiPure}
\hat{A} = P(h_0)\rho_0-P(h_1)\rho_1,
\end{align}
and assigning hypothesis $h_0(h_1)$ if the outcome is one of the positive(negative) eigenvalues of $\hat{A}$.

\begin{figure*}
\includegraphics[trim=0 0 0 0,width=1.9\columnwidth]{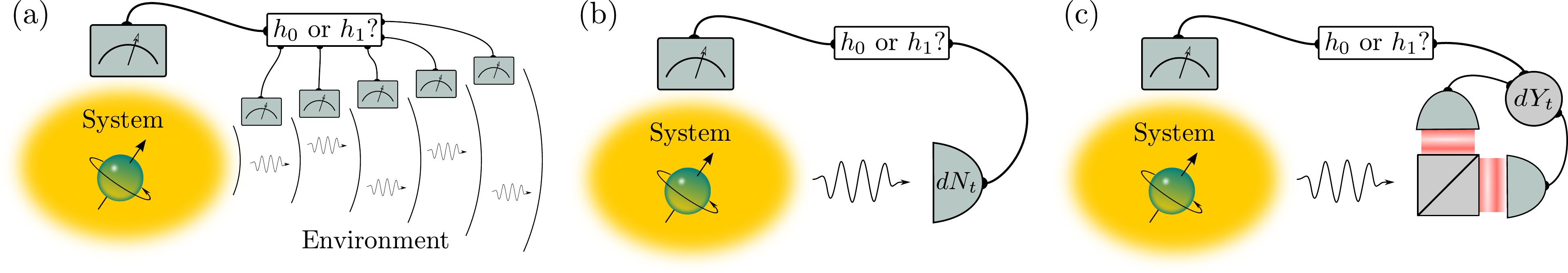}
\caption{
(a) A projective measurement on a system and its environment is performed after they have interacted for a time $T$. Based on the outcome, one of two hypotheses $h_0$ and $h_1$ about their evolution is judged to be more likely than the other.
An experimentally more realistic approach monitors the radiation emitted by the probe system for a time $T$ by (b) photon counting or (c) homodyne detection. The conditional state of the emitter defines the optimal system projection to be performed at the final time $T$ and the most likely hypothesis is inferred from the combined monitoring signal and projection outcome.
}
\label{fig:intro}
\end{figure*}

In this article, we study the use of an open quantum system to distinguish between different Hamiltonian hypotheses. This can, in principle, be accomplished by measuring the optimal observable (\ref{eq:PiPure}) where the $\ket{\psi_i(t)}$ denote the combined states of the system and its environment.
We focus on the common example of a quantum system coupled to a broadband radiation reservoir. A driven system and the quantized radiation field evolves into
entangled states in an infinite dimensional Hilbert space. While these states may differ significantly for different Hamiltonians the, potentially, highly non-local projective measurement (\ref{eq:PiPure}) of combined system and environment observables, see \fref{fig:intro}(a), becomes difficult to achieve in practice. Instead, as illustrated in \fref{fig:intro}, one often has recourse to perform photon counting (b) or field quadrature measurements (e.g. homodyne detection (c)) on the environment.

If the candidate Hamiltonians cause the system to evolve into different steady states, the mean number of emitted photons or the mean homodyne detection signal may be averaged over long enough time to suppress statistical uncertainty about their values such that the Hamiltonian can be inferred with certainty. Faster, and hence more efficient, inference can be made from observation of the correlations in the full noisy measurement record.
To give an example, the steady state yields identical emission rates from atoms excited by one of two strong laser fields, while the time intervals between photon detection events follow distinct oscillatory waiting time distributions.
Optimal inference from any measurement record is obtained by a Bayesian analysis which yields the probabilities $P(h_0|\D)$ and $P(h_1|\D)$ ascribed to each hypothesis based on their prior probabilities and on the full data record $\D$ retrieved until the time $t$ \cite{1355-5111-8-6-002,PhysRevA.64.042105,PhysRevLett.108.170502,PhysRevA.87.032115,PhysRevA.79.022314,PhysRevA.95.022306,PhysRevA.94.032103,PhysRevA.89.052110,PhysRevA.91.012119}.

In this work, we investigate to what extent supplementing continuous monitoring of the emitted radiation from the initial time $t=0$ to a final time $t=T$ by a final projective measurement on the emitter system, allows better distinction between different hypotheses governing the system dynamics.
Due to the measurement backaction associated with continuous monitoring of the environment, the state of the emitter evolves in a conditional manner according to the stochastic measurement signal \cite{QMC}.
In any particular realization of the measurement sequence, the optimal final measurement on the system (\ref{eq:PiPure}) is thus conditioned on the detection record $\DT$ obtained up until the final time $T$,
\begin{align}\label{eq:PiConditional}
\hat{A}^\DT = P(h_0|\DT)\rho^{(\DT)}_0(T)-P(h_0|\DT)\rho^{(\DT)}_1(T).
\end{align}
Here the information extracted from the environment is incorporated in the conditional candidate states $\rho_i^{(D_T)}(T)$ and their probabilities
updated by Bayes rule, $P(h_i) \rightarrow P(h_i|\DT)$.

The continuous monitoring and conditioned evolution of quantum states have for instance been realized in experiments with superconducting qubits \cite{murch2013observing,PhysRevA.96.022104,PhysRevX.6.011002} and optomechanical systems \cite{PhysRevLett.114.223601}.
After homodyne or heterodyne detection of the radiation signal has been performed until time $T$ on for example a super conducting qubit, a final system projection can be achieved in these experiments by applying a strong, dispersively coupled probe field \cite{murch2013observing,PhysRevLett.114.090403}.
We compare such realistic measurement strategies with the theoretical limit for distinguishing different hypotheses.
See also \cite{Jacobs2007FeedbackCF} for an alternative, adaptive approach to hypothesis testing and state discrimination with continuous measurements.

The article is organized as follows.
In Section~\ref{sec:2} we outline the main ideas of hypothesis testing with monitored quantum systems and we recall a lower (quantum) bound for the error probability.
In Section~\ref{sec:3} we present numerical simulations which illustrate and exemplify different aspects of our theory.
In Section~\ref{sec:4} we provide a conclusion and an outlook.

\section{Bayesian analysis of a measurement record}\label{sec:2}

We consider a system subject to a sequence of measurements or continuous monitoring from time $t=0$ to a final time $t=T$.
During this phase, a signal $\dD$ is recorded and by $\D$
we denote the full signal obtained between time $0$ and $t$. Under hypothesis $h_i$ any given realization of $\D$ has a probability $P(D_t|h_i)$ determined from the conditional candidate quantum state $\rhoc_i(t)$.
Bayes rule yields the corresponding update of the likelihood $P(h_i|D_t)$ assigned to each hypothesis,
\begin{align}\label{eq:BayesMonitor}
P(h_i|\D) = \frac{P(\D|h_i)P(h_i)}{\sum_j P(\D|h_j)P(h_j)}.
\end{align}

The backaction of the measurement associated with the outcome $\dD$ applies directly on the current state of the system, $\rhoc(t) \rightarrow {\hat{M}(\dD)\rhoc(t)\hat{M}^\dagger(\dD)}/{P(\dD)}$. Here the sum (integral) of the positive-operator valued measure (POVM) over all possible detection outcomes yields the identity, $\sum_{dD_t} \hat{M}^\dagger(\dD)\hat{M}(\dD) = I$.
The POVM formalism includes both projective measurements, in which case the $\hat{M}(\dD)$ denote projection operators, as well as more
general measurements, involving, e.g., projective measurements on ancilla systems after they have interacted with the system. Between measurements, the system evolves subject to the Hamiltonian that we want to discriminate.


If the state is not renormalized after application of the POVM backaction operators, we retain the evolution of an unnormalized state $\rhotc(t)$,
\begin{align}\label{eq:POVMunnormlized}
\rhotc(t) \rightarrow  \hat{M}(\dD)\rhotc(t)\hat{M}^\dagger(\dD),
\end{align}
whose reduction in norm is just the probability to obtain the signal $\dD$. This implies that at the final time $T$, the probability $P(dD_T)\cdots P(dD_{2dt})P(dD_{dt})P(dD_0)$
for the full signal $\DT$ is given by the trace of $\rhotc(T)$. Hence, by evolving the unnormalized state under each of the two candidate hypotheses conditioned on the signal \textit{actually} recorded in a given experiment, one may by \eqref{eq:BayesMonitor} obtain the relative likelihoods of each hypothesis as $P(h_i|\D)\propto \mathrm{Tr}(\rhotc_i(t))$. Since any specific trajectory for $\D$ is very unlikely, $\mathrm{Tr}(\rhotc_i(T))$ becomes very small even for the true hypothesis and for numerical purposes it is favorable to propagate instead the log-likelihood $\log[P(h_i|\D)]$. See \cite{PhysRevA.87.032115} for a detailed account of Bayesian inference with continuously monitored quantum systems.

In the next subsection we specialize to cases, where the measurements are carried out continuously in time on the radiation field emitted by the quantum system of interest. The two generic setups of counting-type measurements with discrete detection events and diffusion-type measurements with continuous but infinitesimal backaction are discussed, and \eqref{eq:POVMunnormlized} is replaced by stochastic master equations, suitable for numerical propagation of $\rhotc(t)$. For simplicity we assume that there is only a single decay channel but the expressions may readily be generalized to multi-channel cases and alternative environmental couplings.

\subsection{Photon counting and homodyne detection}
In \fref{fig:intro}(b), the florescence from the probe system is detected by a photon counter with quantum efficiency $0\leq\eta\leq1$ and the photon counting signal $N_t$ until time $t$ constitutes the detection record $D_t$.
During each short time interval $dt$ there are two possible detection outcomes: no photon $dN_t = 0$ or one photon $dN_t = 1$, where $P(dN_t = 1)= \eta\Tr{\Cd\C\rhon(t)}dt$ is given by the (normalized) state $\rhon(t)$ of the system. Here $\C=\sqrt{\gamma}|g\rangle \langle e|$ denotes the quantum jump operator from an excited $|e\rangle$ to a lower state  $|g\rangle$.

The conditional evolution of the unnormalized state, in turn, obeys a linear stochastic master equation \cite{PhysRevLett.68.580,QMC},
\begin{align}\label{eq:meCount}
\begin{split}
d\rhotn =& \left(\mathcal{K}dt+ \mathcal{B}dN_t\right)\rhotn,
\end{split}
\end{align}
where $\mathcal{K}\rho = -i[\H,\rho] +(1-\eta)\C\rho\Cd-\frac{1}{2}\{\Cd\C,\rho\}$ and $\mathcal{B}\rho = \eta\left(\C \rho\Cd-\rho\right)$.

As depicted in \fref{fig:intro}(c), a homodyne detector mixes the florescence with a strong local oscillator field on a beam splitter, and the signal $dY_t$ is obtained as the intensity difference between the two output ports.
Homodyne detection is sensitive to the phase of the emitted radiation which may be favorable when probing certain dynamics of the system.
The recorded signal $dY_t$ in each short time interval $dt$ has a mean value determined by the current state $\rhoy(t)$ of the system,
\begin{align}\label{eq:dY}
dY_t = \Tr{\mathcal{X}_\Phi\rhoy(t)}dt+dW_t.
\end{align}
with $\mathcal{X}_\Phi\rho =\sqrt{\eta}\left(\C \mathrm{e}^{-i\Phi}\rho+\rho\Cd \mathrm{e}^{i\Phi}\right)$ where $\Phi$ is the phase of the local oscillator.
Random, white-noise fluctuations around the mean are represented by infinitesimal Wiener increments which are uncorrelated, normal distributed stochastic elements with zero mean and variance $dt$.
Since the signal depends only weakly on the state of the system, the backaction associated with homodyne detection is infinitesimal and \eqref{eq:POVMunnormlized} is equivalent to a diffusion type linear stochastic master equation for the conditional evolution of the unnormalized state \cite{QMC},
\begin{align}\label{eq:meHomo}
d\rhoty= \left(\mathcal{L} dt+\mathcal{\mathcal{X}}_\Phi dY_t  \right) \rhoty,
\end{align}
where $\mathcal{L}\rho = -i[\H,\rho] +\C\rho\Cd -\frac{1}{2}\{\Cd\C,\rho\}$.

Upon acquiring a measurement signal,
the relevant stochastic master equation, (\ref{eq:meCount}) or (\ref{eq:meHomo}), may be solved for each hypothesis.
The corresponding candidate states are all initialized in the (known) initial state of the system, but normalized to the prior probabilities assigned the particular hypothesis, $\mathrm{Tr}(\rho_i^{D_0}(t=0)) = P(h_i)$. This way the evolving likelihood distribution over the possible hypotheses is directly given by the traces of the corresponding conditioned density matrices, $\tilde{\rho}_0^{D_t}(t),\ \tilde{\rho}_1^{D_t}(t)$.

\begin{figure}
\includegraphics[trim=0 0 0 0,width=0.95\columnwidth,left]{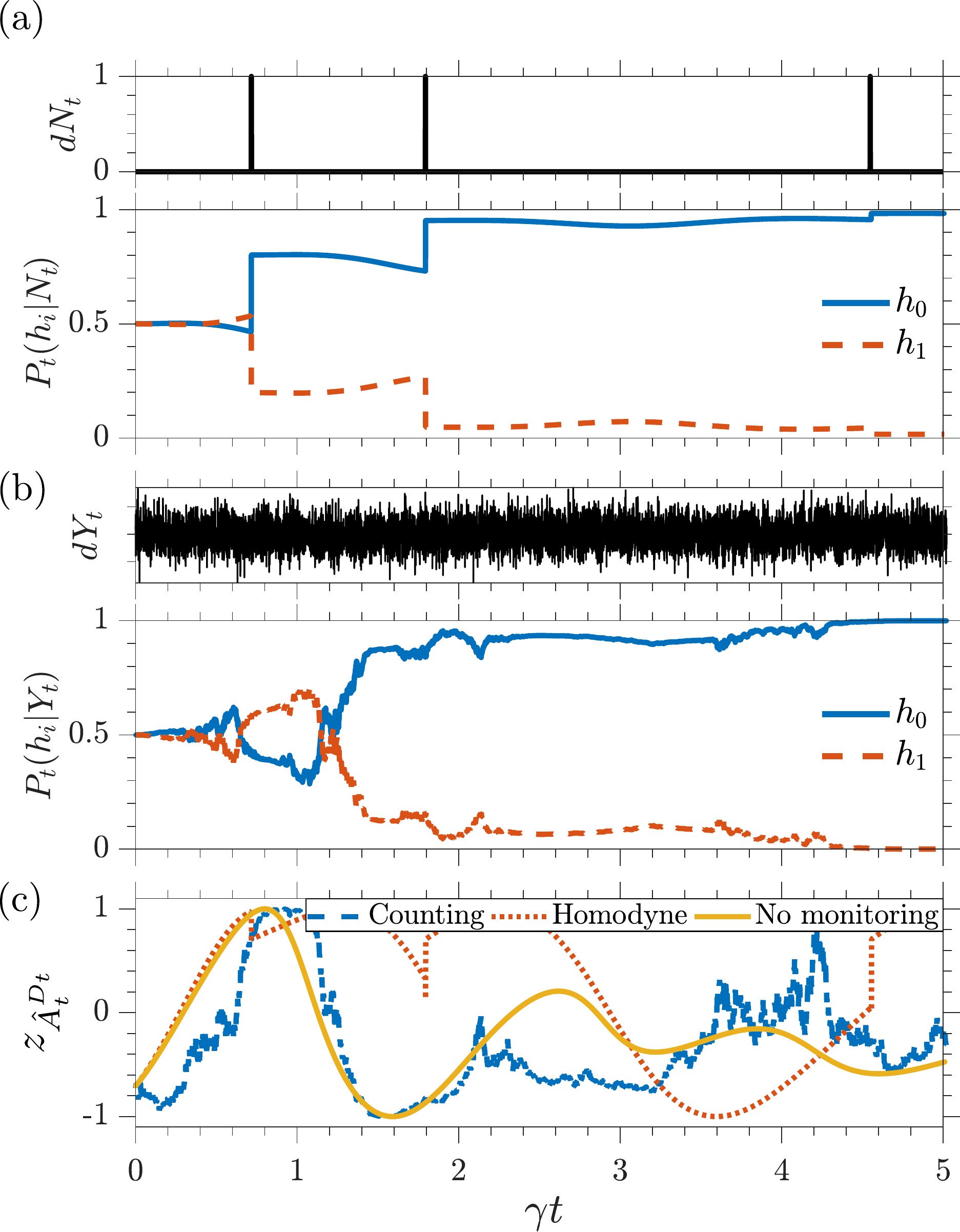}
\caption{
Simulated monitoring of a driven two-level system by (a) photon counting and (b) homodyne detection with the purpose of discriminating two hypotheses $h_0$ ($\Omega_0 = 2\gamma$) and $h_1$ ($\Omega_1 = 4\gamma$) for the Rabi frequency.
The simulations are made assuming $h_0$ to be the true hypothesis and with a detector efficiency $\eta=1$ and in (b) a local oscillator phase $\Phi = -\pi/2$.
The second and fourth panels show the evolution of the probabilities (\ref{eq:BayesMonitor}) for the two hypotheses  conditioned on (a) the photon counting signal and (b) the noisy homodyne current shown in the first and third panels.
The lower panel (c),  shows the $z$-component $z_{\hat{A}_t^{\D}} \propto \tr{\sz\hat{A}_t^{\D}}$ of the optimal Pauli measurement observable (clarified in the main text) if monitoring is stopped at any given time. We observe that this optimal system measurement differs for the three cases of counting, homodyne detection and unobserved, dissipative emitter dynamics.
}
\label{fig:homodyneCountingComparison}
\end{figure}

To illustrate the Bayesian inference protocol, we simulate in \fref{fig:homodyneCountingComparison} perfect monitoring of a two-level system with the purpose of discriminating two hypotheses for the resonant driving with a Rabi frequency of either
$\Omega_0$ or $\Omega_1$. I.e, we test the two Hamiltonian hypotheses:
$
\H_0 =  \frac{\hbar\Omega_0}{2}\sx
$
and
$
\H_1 =  \frac{\hbar\Omega_1}{2}\sx
$.
The signals, $dN_t$ from photon counting and $dY_t$ from homodyne detection, in the upper panels of (a) and (b) are sampled from the true hypothesis which we assume to be $h_0$. Conditioned on these signals, the (unnormalized) candidate states $\rhotc_i(t)$ with $\D=N_t,Y_t$ evolve according to Eqs.~(\ref{eq:meCount})~and~(\ref{eq:meHomo}), respectively. Their traces and the condition $P(h_0|\D)+P(h_1|\D)=1$ yield the time evolution of the inferred probabilities for each hypothesis as shown in the lower panels of (a) and (b). We assume equal priors $P(h_0)= P(h_1) = 1/2$.

For photon counting in (a) the probability updates are dominated by three photo detection events while periods with no detections lead to a less pronounced, continuous update.
The noisy homodyne signal in (b), on the other hand, holds only very little information in each individual time-bin and here the probabilities continuously converge to reveal the true hypothesis.
In both cases, at the final time $t=5\gamma^{-1}$ the accumulated signals are seen to favor the true hypothesis ($h_0$) with almost unit probability.
A figure of merit for a particular measurement strategy is the speed at which we arrive at perfect distinction.

\subsection{Supplementing continuous monitoring by a projective measurement}
If the hypotheses are not sufficiently discriminated at the end of the probing at time $T$, it may be possible to extract further information by a direct measurement on the emitter system. Due to the continuous monitoring, the emitter is assigned the conditional candidate states $\rho^{(\DT)}_i(T)$, while the probabilities that we ascribe to these states, $P(h_i|\DT)$  are given by the traces of the unnormalized density matrices.

The optimal projective measurement we can perform on the system then concerns the system observable $\hat{A}^{D_T}_T$ defined in \eqref{eq:PiConditional}.
For a two-level system, the projective measurement of any observable $\hat{A}$ is equivalent to the measurement of a Pauli spin component along a specific unit vector $(x_A,y_A,z_A)$ with $u_A\propto \tr{\hat{\sigma}_u A}$.
In \fref{fig:homodyneCountingComparison}(c) we visualize the optimum observable $\hat{A}^{D_T}_T$ if the continuous monitoring, yielding the signals in the upper panels of (a) and (b), is terminated at the corresponding point in time. In this example the unit vector, designating the direction of the spin measurement, is confined to the $(y,z)$-plane and we show its $z$-component

During each experimental realization, $\hat{A}_t^{\D}$ assumes a stochastic value, which is different from the one that optimally discriminates the states of an unobserved system governed by the corresponding Lindblad master equation $d\rho/dt = \mathcal{L}\rho$.
With homodyne detection the measurement observable, represented by the blue noisy trace in \fref{fig:homodyneCountingComparison}(c), is seen to fluctuate around the full, yellow curve,  pertaining to the unmonitored system, while with photon counting, large deviations arise accompany the quantum jumps of the system state.

The possible eigenvalues $\lambda$ of the measurement observable $\hat{A}^{D_T}_T$ occur under hypothesis $h_i$ with probability $P(\lambda|h_i)=\Tr{\Pi_\lambda \rho_i(t)}$, where $\Pi_\lambda$ is the projector on the affiliated eigenstate of the operator $\hat{A}^{D_T}_T$.
According to Bayes rule the combined information from the monitoring and from the system projection hence leads to an update of the probabilities assigned to each hypothesis
\begin{align}\label{eq:assignProb}
P(h_i|\DT,\lambda) = \frac{P(\lambda|h_i)P(h_i|\DT)}{P(\lambda)}.
\end{align}
The hypothesis $h_m$ with the largest likelihood $P(h_m|\DT,\lambda)$ is the preferred one, and averaged over many independent realizations of the final projective measurement, the fraction of erroneous assignments based on that choice will be given by the generalization of \eqref{eq:QeMinPure} to mixed states,
\begin{align}\label{eq:QeMin}
\Qe = \frac{1}{2}\left[1-\left|P(h_0|\DT)\rho^{(\DT)}_0(T) -P(h_1|\DT) \rho^{(\DT)}_1(T)\right|\right],
\end{align}
where $|O| \equiv \Tr{\sqrt{O^\dagger O}}$. To obtain the error probability of a given measurement scheme, we however still need to numerically evaluate the conditional states and probabilities and average \eqref{eq:QeMin} over the random outcomes of the continuous monitoring.

Note that \eqref{eq:QeMin} can also be applied to the distinction of mixed states or of the (unconditioned) candidate density matrices of a system evolving under different Hamiltonian hypotheses and leaking into an un-monitored environment. A recent comparison of probing by measurements on a system alone and on both a system and its environment shows the ability of the latter to better exploit (initial) entanglement among its sub-components  \cite{albarelli2018restoring}.

\subsection{The quantum bound}\label{sec:Qbound}
The minimum achievable error associated with any hypothetical detection of the radiation emitted by a system and a final detection on that system itself is determined by our ability to discriminate the pure states of the combined system and environment, resulting from the different Hamiltonian hypotheses. These (un-monitored) states are themselves intractable by numerical means, but if the Born-Markov approximation applies for the radiative emission process, their quantum overlap can be evaluated as the trace of an effective density matrix $\rzo(t)$ acting only on the state space of the emitter system:
$\langle \psi_0(t)|\psi_1(t)\rangle = \Tr{\rzo(t)}$.
This matrix evolves from the initial pure state of the system according
to the following master equation \cite{PhysRevLett.114.040401,kiilerich2018multi},
\begin{align}\label{eq:2sided}
\begin{split}
\frac{d\rzo}{dt} = &-i\left(\H_0\rzo-\rzo \H_1\right)
\\ &+
\sum_j\left[\cjz\rzo\cjo^\dagger-\frac{1}{2}\left(\cjz^\dagger\cjz\rzo+\rzo\cjo^\dagger\cjo\right)\right].
\end{split}
\end{align}
Note that the matrix evolves under the action from the left and right with the different candidate Hamiltonians and with different relaxation operators, $\cjz$ and $\cjo$, representing cases where the hypotheses concern the damping of the system.
Unlike the conventional Lindblad master equation, \eqref{eq:2sided} does not preserve the trace, and the overlap between candidates for the full system and environment quantum states attains non-trivial values, resulting in a time dependent value of $\Qe^{(\text{min})}(t)$ as given in \eqref{eq:QeMinPure}.
This quantity represents a lower (quantum) bound on the probability of assigning a false hypothesis based on \textit{any} combined quantum measurement performed on the environment in the time interval $[0,t]$ and on the emitter system at the time $t$, corresponding to the situation depicted in \fref{fig:intro}(a). In the next section we compare the achievements of testing using continuous measurements and Bayesian discrimination with this minimum.

\section{Numerical investigations}\label{sec:3}

\subsection{Error probabilities under different detection models}
\label{sec:EP}
\begin{figure}
\includegraphics[trim=0 0 0 0,width=0.95\columnwidth,left]{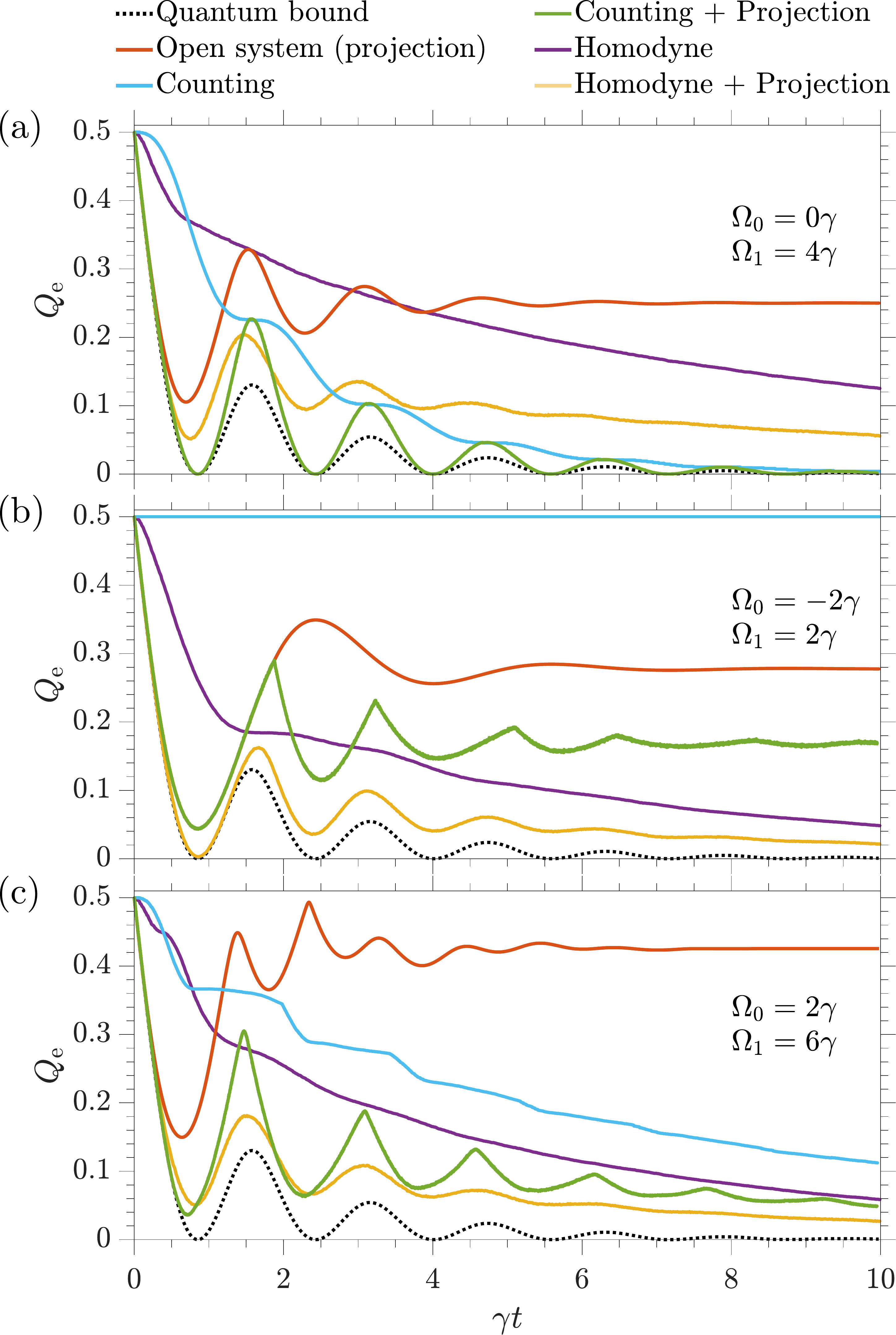}
\caption{Temporal evolution of the error probability in assigning one of two hypotheses $\Omega_0$ and $\Omega_1$ for the Rabi driving frequency of a two-level system.
The three plots correspond to different pairs of Rabi frequency candidates as annotated in the figure windows and the system is prepared in the ground state at $t=0$.
Results are shown for each of the different measurement schemes discussed in this paper. The error probabilities pertaining to monitoring protocols with perfect detection $\eta=1$ are sampled from $M=100.000$ simulations (see main text).
}
\label{fig:QeHomodyneCounting}
\end{figure}

To address the performance of the different monitoring schemes, we turn to the associated error probability $\Qe$. We consider both the case where the probability update is based solely on the detection signal, \eqref{eq:BayesMonitor}, and the case where the signal is combined with a final optimized projective measurement on the system, \eqref{eq:assignProb}.

The probabilities pertain to the average over many independent experimental realizations. However, they are non-linear functionals of the conditional states so there is no deterministic theory which allows their evaluation.
Instead we have recourse to perform a large number $M$ of simulations of the full measurement sequence and Bayesian inference.
We repeated the simulations assuming each of the two hypotheses $h_0$ and $h_1$ to be true.
In testing based on the detection signal $\DT$ alone, hypothesis $h_i$ is assigned if
$P(h_i|\DT)>1/2$.
The probability in \eqref{eq:Qe} to discard a true hypothesis $h_j$ is then estimated by
$
P(m=i|h_j) = n_{i}^{(j)}/M,
$
where $n_{i}^{(j)}$ is the number of samples assigning $h_i$ when $h_j$ is true.
When a final system projection with outcome $\lambda$ is included in the procedure, the assignment is dictated by $P(h_i|\DT,\lambda)>1/2$ and the error probability is given directly by \eqref{eq:QeMin}.

The resulting error probabilities for our two-level example are compared to the quantum bound and to that of a projective measurement on the open system alone in \fref{fig:QeHomodyneCounting}.
Curves are shown for three pairs of Rabi frequency candidates. They are all separated by $\Omega_1-\Omega_0 = 4\gamma$, and therefore the error probabilities have the same quantum lower bound \cite{PhysRevLett.114.040401}, but their particular offsets make either counting or homodyne detection more advantageous.
All protocols yield larger error probabilities than the quantum bound. This means that none of the measurement strategies are optimal in the sense that they are able to extract all information from the full state of the system and its environment.

A photon counting signal is sensitive to the intensity of the emitted radiation and hence reflects the excitation of the two-level system. As seen in (a) this makes it near ideal to distinguish $\Omega_0 = 0$, which leads to no photon emissions, from a strong drive $\Omega_1=4\gamma$. 
The counting signal alone generates a much smaller error probability than the homodyne signal and approaches zero on a timescale similar to that of the quantum bound. When combining the counting signal with a final system projection, the error probability follows the quantum bound closely at short times and it shows that we may at specific finite probing times distinguish the hypotheses with certainty. These are points in time where the non-zero Rabi frequency $\Omega_1$ assures an atomic or a photonic excitation.

The photon count is, however, insensitive to the phase of the emitted radiation and to the coherences in the two-level system. As a consequence, the two candidates $\Omega_0=-2\gamma$ and $\Omega_1 = 2\gamma$ in (b) can not be discriminated by the photon counting signal alone; i.e. $\Qe(t)=1/2$ for all times. Homodyne detection is, on the other hand, highly sensitive to the phase of the emitted radiation and when combined with a final system projection, the associated error matches the quantum bound for $\gamma t\lesssim 1.5$ after which it remains close to the bound.

Since for the case studied in (b) the photon count alone holds no discriminatory power, one might expect
that supplementing a counting signal with a final system projection yields
an error probability identical to that pertaining to a projective measurement on the mixed state of an unmonitored system. Nevertheless, it is seen than for $\gamma t\gtrsim 1.75$, counting the photo emissions reduces the final error probability by around $10\%$.
This illustrates an additional advantage of monitoring the environment. Subject to backaction, the system state remains pure and experiences a transient behavior which generally depends more strongly on the particular hypothesis than the mixed state of the unmonitored system. This allows more information to be extracted from the final system measurement. Previous works identify similar mechanisms at play in parameter estimation with monitored systems \cite{PhysRevA.87.032115,PhysRevA.89.052110,PhysRevA.91.012119,PhysRevA.94.032103,albarelli2018restoring}.

The candidate values $\Omega_0=2\gamma,\ \Omega_1=6\gamma$ in (c) can be distinguished both by the excitation and the coherence of the system. It is evident that while homodyne detection is slightly better than counting for these particular values, they both perform well and reach within $5-10\%$ of the quantum bound.

\subsection{Finite detector efficiency}
While the simulations in Figures~\ref{fig:homodyneCountingComparison}~and~\ref{fig:QeHomodyneCounting} assume perfect monitoring, any real experiment suffers from finite detection efficiency $\eta<1$.
If the environment is monitored with perfect efficiency $\eta=1$, the system state remains pure but if, e.g., some photo emissions are missed by the detector we are unable to perfectly track the state of the system and the conditional state $\rhoc(t)$ evolves to a statistical mixture.
Consequently, in addition to the direct decrease in information available from the monitoring signal, the final system measurement is performed on a mixed state with, in general, less discriminatory power.
\begin{figure}
\includegraphics[trim=0 0 0 0,width=1\columnwidth,left]{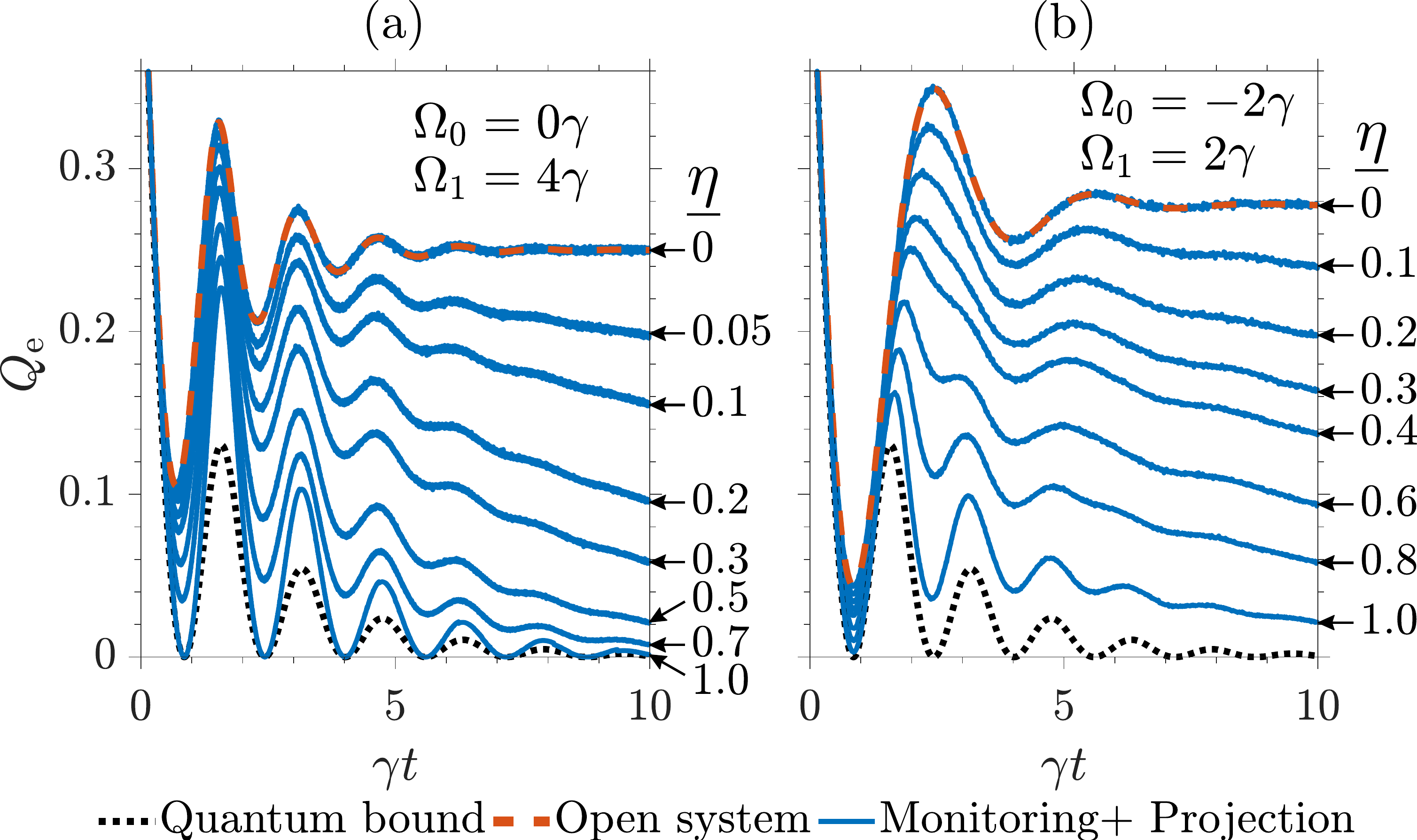}
\caption{
Temporal evolution of the error probability in assigning one of two hypotheses $\Omega_0$ and $\Omega_1$ for the Rabi driving frequency of a two-level system.
The candidate values are annotated in the figure windows and the system is prepared in the ground state at $t=0$.
The full, blue curves, concerning monitoring by photo detection (a) and by homodyne detection (b) combined with a final system projection, are sampled from $M=100.000$ simulations (see main text) with different values of the detection efficiency $\eta$ as indicated on the right hand side of each plot.
For comparison, we show also the quantum bound (dotted curve) and error probability associated with a projective measurement on an open system (dashed, red curve).
}
\label{fig:etaDependence}
\end{figure}

To probe these effects, we show in \fref{fig:etaDependence} the (sampled) error probability for different values of $\eta$. For (a) photon counting we focus on the candidates $\Omega_0=0,\, \Omega_1 = 4\gamma$ and for (b) homodyne detection $\Omega_0=-2\gamma,\, \Omega_1 = 2\gamma$ where each of the two methods work particularly well.
As $\eta$ decreases, the error probability $\Qe(t)$ undergoes a smooth transition from the perfect detection case studied in \fref{fig:QeHomodyneCounting} to the case of a projection measurement performed on the mixed state of the system alone
in the limit $\eta\rightarrow 0$.
For the parameters used in this example, the photon counting protocol in (a) is surprisingly robust to detector imperfections. This is due to the fact, that as explained in Section~\ref{sec:EP}, even a single photo detection completely rules out the hypothesis $\Omega_0=0$.
While the homodyne example in (b) shows a more linear increase in the error probability as the detector efficiency deteriorates, both plots demonstrate that even with fairly large imperfections, monitoring the environment severely
improves the hypothesis testing capabilities of an open quantum system.
This is due to the fact that the monitoring induces transient evolution in the system which depends more strongly on the system parameters than the steady state.

\section{Conclusion and outlook}
\label{sec:4}
We have investigated how hypothesis testing with an open quantum system may be improved by monitoring the radiative environment to which it is coupled.
We propose to supplement the information retrieved directly from the monitoring signal with a final system measurement optimized according to the conditional state.
For reasons of clarity, we restricted our attention to just two distinct hypotheses, but the Bayesian analysis is readily generalized to cases with multiple candidates and in Ref.~\cite{kiilerich2018multi} we present an efficient numerical approach to evaluate the quantum bound and define the optimal system projection when multiple hypotheses are in play.

It was found that, while monitoring by a photon counter or a homodyne demodulator allows the extraction of much of the information leaked from the open system into the field, the error probability in these schemes does not reach the fundamental quantum bound.
As explained in the introductory section~\ref{sec:1}, this is not surprising since generally the optimal measurement is highly non-local on the full system and environment.

\begin{figure}
\includegraphics[trim=0 0 0 0,width=0.95\columnwidth]{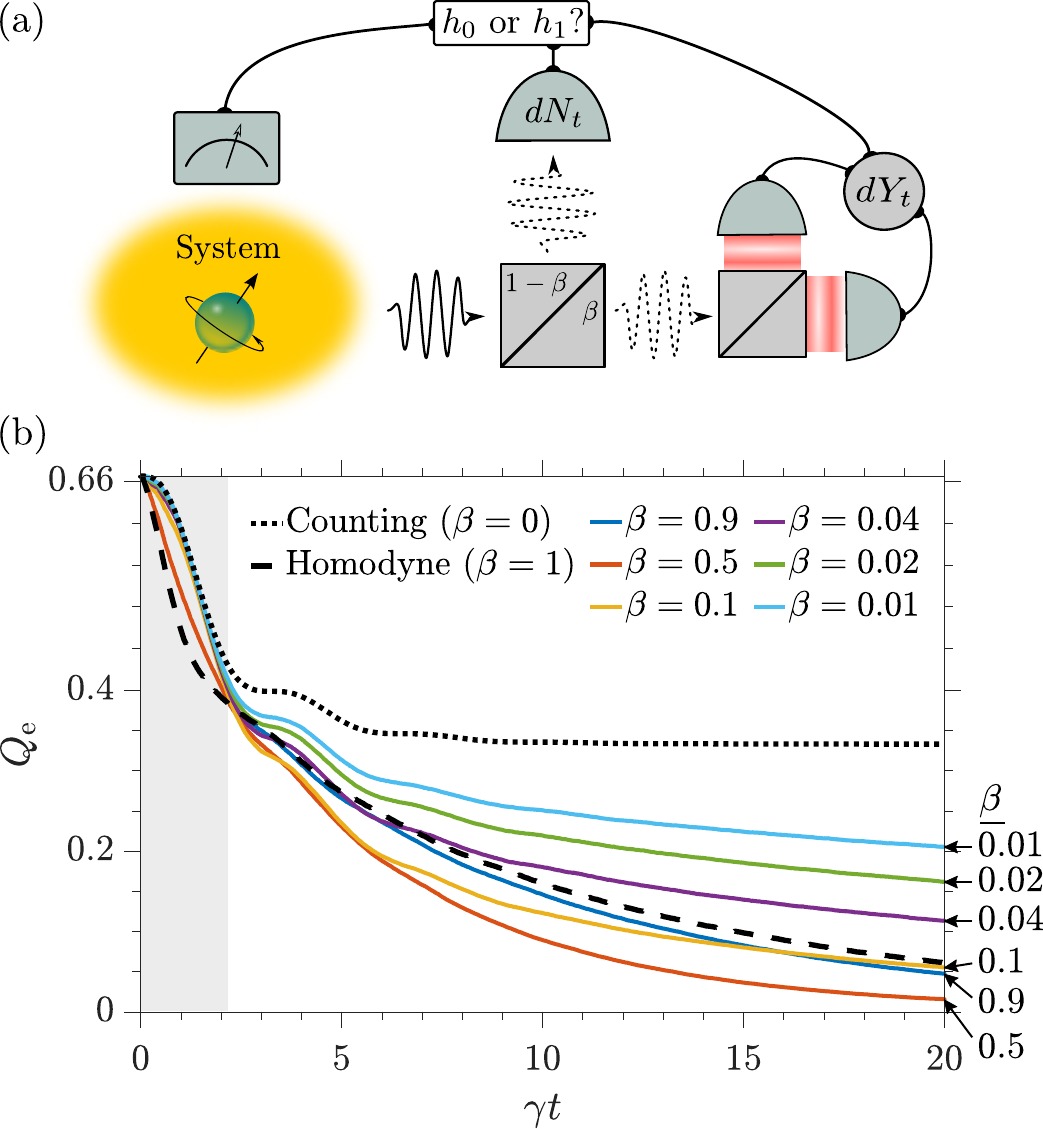}
\caption{
(a)
A fraction $\beta$ of the radiation emitted by a probe system is collected by a homodyne demodulator while the remaining fraction $1-\beta$ is directed to a photon counter. The system state, which defines the optimal system projection to perform at the final time $T$, is conditioned on both the photon count and the homodyne signal.
(b) Temporal evolution of the error probability in assigning one of three hypotheses $\Omega=0,\pm 2 \gamma$ for the Rabi frequency of a driven two-level system based on the two monitoring signals, $N_t$ and $Y_t$ of the hybrid monitoring scheme in (a). The cases of pure counting ($\beta=0$) and pure homodyne detection ($\beta=1$) are compared to different hybrid schemes with $0 < \beta < 1$. Pure homodyne detection is only optimal for times $\gamma t\lesssim 2.3$ (shaded area).
The error probabilities are sampled from $M=100.000$ simulations.
 }
\label{fig:hybrid}
\end{figure}

From the results presented in \fref{fig:QeHomodyneCounting}, it is clear that homodyne detection and photon counting yield different reductions in the error probability at different stages in the evolution. I.e., at some points in time either homodyne detection or photon counting is more efficient than the other.
To allow both possibilities in a single experiment, the setup illustrated in \fref{fig:hybrid} splits the
radiation emitted by the system such that a fraction $1-\beta$ is monitored by a photon counter and the remaining $\beta$ fraction is subject to homodyne detection.
The conditional, unnormalized state $\tilde{\rho}^{(N_t,Y_t)}(t)$ then evolves according to both monitoring signals,
\begin{align}\label{eq:meCountHom}
\begin{split}
d\tilde{\rho}^{(N_t,Y_t)} &= \Big(\left[(1-\beta)\mathcal{K}+\beta\mathcal{L}\right]dt
\\
&+(1-\beta)\mathcal{B}dN_t
+\sqrt{\beta}\mathcal{X}_\Phi dY_t
\Big)\tilde{\rho}^{(N_t,Y_t)}.
\end{split}
\end{align}
A similar scheme applies the homodyne setup, \fref{fig:intro}(c) but with a local oscillator of variable strength $\xi$ \cite{zhang2012mapping}.
Conventional homodyne detection is realized in the limit of large $\xi$, while with a weak local oscillator the setup effectively counts photons.

The significance of such \textit{hybrid} schemes is more apparent in scenarios with multiple distinct hypotheses, and in \fref{fig:hybrid}(b) we illustrate this by considering the differentiation of three discrete values $\Omega=0,\pm 2\gamma$ of the Rabi frequency in our two-level model. For sake of argument, we consider only monitoring without a final system projection.
As discussed in Section~\ref{sec:EP}, pure photo detection $(\beta=0)$ is only sensitive to the absolute value of $\Omega$, and hence the error probability never reaches values lower than $Q_\mathrm{e}=1/3$, signifying perfect discrimination between $\Omega=0$ and the values $\pm 2 \gamma$ which are, on the contrary, indistinguishable.
When even a small fraction $\beta>0$ of the intensity of the emission signal is monitored by a homodyne demodulator, however, the combined signal is able to perfectly distinguish the three hypotheses if sufficient time is alloted.
Interestingly, while pure homodyne detection ($\beta=1$) is optimal for times $\gamma t\leq 2.3$ (shaded area), hybrid schemes with $0<\beta<0.9$ converge faster to perfect discrimination because a photon counting signal very efficiently discriminates $\Omega=0$ from any non-zero values.
Notice, finally, the large reduction in the error probability from the $\beta=0$ to the $ \beta=0.01$ case. This is because just $1\%$ of the intensity amounts to $10\%$ of the amplitude, which is the relevant observable in homodyne detection, and  leaves the counting signal virtually unaltered.

By using a beamsplitter with a tunable transmittance $\beta(t)$ or by adjusting the local oscillator strength $\xi(t)$, the effective monitoring scheme can be updated in a time dependent manner in order to further optimize the information extracted at each point in time.
Such a task may be guided by intuition or achieved by numerical optimal control based on the formalism presented in this article.

\section{Acknowledgements}
The authors would like to thank Peng Xu for helpful discussions and acknowledge financial support from the Villum Foundation.
A.\,H.\,K. further acknowledges financial support from the Danish Ministry of Higher Education and Science.

%

\end{document}